# 14

# Navigating Turbulence

## The Challenge of Inclusive Innovation in the U.S.–China AI Race

*Jyh-An Lee and Jingwen Liu*

## Introduction

Artificial intelligence (AI) has not only transformed the global economy but also redefined the international power hierarchy. The current "AI arms race" is often likened to the space race of the 1960s or the nuclear arms race from the 1950s to the 1980s.[1] Consequently, it is hardly surprising that some assert that the dominant force in AI technology may ultimately govern global leadership.[2] Although AI development is making breakthroughs around the world, many believe that global AI competition ultimately boils down to a contest between the United States and China.[3]

The AI market in the United States continues to drive significant venture capital investment. According to data from Crunchbase, AI startups secured US$18.9 billion in the third quarter of 2024, accounting for 28 percent of all venture funding during that period.[4] The strong culture of entrepreneurship and a well-established practice in technology investment have provided U.S. AI firms with an unparalleled advantage in the global arena. OpenAI's launch of ChatGPT in November 2022 has further showcased to the world the extraordinary potential of generative AI. Beyond ChatGPT, American companies have developed most of the world's advanced large-language

---

[1] Lee Kai-Fu, AI Superpowers China, Silicon Valley and the New World Order (2018).

[2] Radina Gigova, *Who Vladimir Putin Thinks Will Rule the World*, CNN (Sept. 1, 2017), https://edition.cnn.com/2017/09/01/world/putin-artificial-intelligence-will-rule-world/index.html.

[3] *See, e.g.*, Neil Savage, *The Race to the Top Among the World's Leaders in Artificial Intelligence*, Nature Index (Dec. 9, 2020), https://www.nature.com/articles/d41586-020-03409-8.

[4] Gené Teare, *Global Funding Slowed in Q3, Even as AI Continued to Lead*, Crunchbase News (Oct. 3, 2024), https://news.crunchbase.com/venture/global-startup-funding-recap-q3-2024/.

models (LLMs), including Claude from Anthropic, BERT, Gemini, and Gemma from Google, Llama from Meta, and Grok from xAI.

However, AI companies in Silicon Valley are now encountering fierce competition from their counterparts on the other side of the globe. In recent years, China's AI industry has witnessed rapid growth. According to the China Center for Information Industry Development, the country's core AI industry has grown to an approximate scale of US$69.48 billion in 2023, encompassing over 4,400 enterprises.[5] The market scale for AI large models is projected to reach US$2.1 billion, marking a 110-percent year-on-year increase.[6] On March 16, 2023, Chinese technology giant Baidu launched its ERNIE Bot, a new-generation knowledge-enhanced LLM. The latest iteration, released on June 21, 2023, was reported to have surpassed ChatGPT-4 in several Chinese-language capabilities.[7]

The introduction of DeepSeek in early 2025 has shed further light on the current capabilities of Chinese AI technology.[8] Despite the United States' heightened export controls on semiconductor chips to constrain China's AI advancements, DeepSeek-R1 has demonstrated performance metrics in mathematics, coding, and complex reasoning that are on par with OpenAI's leading models. For instance, in the realm of mathematics, DeepSeek-R1 achieved a score of 79.8 percent on the AIME (Artificial Intelligence Math Evaluation) 2024 benchmark, slightly surpassing OpenAI o1's 79.2 percent. Similarly, in coding proficiency, DeepSeek-R1 excelled at the 96.3rd percentile on Codeforces, closely trailing OpenAI o1's 96.6th percentile.[9]

In the midst of a fervent global arms race in technology, governments have played pivotal roles in fostering AI innovation. Strategic plans on the use and deployment of AI can be found across various nations.[10] In the United States, strategic AI policies started with the Obama administration's publication of the report on "Preparing for the Future of Artificial Intelligence" and the National Artificial Intelligence Research and Development Plan in 2016.[11] This trajectory continued under President Trump, who launched the American AI Initiative to preserve U.S. leadership in AI in his first administration.[12] Further advancing these efforts, the Biden administration, in May 2023, revised the National Artificial Intelligence Research and Development Strategic Plan, building upon its predecessors from 2016 and 2019.[13] This update introduced a principled, coordinated strategy for international collaboration in AI research, enhancing eight strategic directions.[14]

---

[5] *China to Launch AI Plus Initiative: Government Work Report*, GLOB. TIMES (Mar. 5, 2024), https://www.globaltimes.cn/page/202403/1308210.shtml.

[6] *Id.*

[7] *Introducing ERNIE 3.5: Baidu's Knowledge-Enhanced Foundation Model Takes a Giant Leap Forward*, BAIDU RSCH. (June 27, 2023), https://research.baidu.com/Blog/index-view?id=185.

[8] Wang Fan & João da Silva, *DeepSeek: How China's "AI Heroes" Overcame US Curbs to Stun Silicon Valley*, BBC (Jan. 28, 2025), https://www.bbc.com/news/articles/czepw096wy4o.

[9] Craig S. Smith, *DeepSeek: How China's AI Innovators Are Challenging the Status Quo*, FORBES (Jan. 22, 2025), https://www.forbes.com/sites/craigsmith/2025/01/22/deepseek-how-chinas-ai-innovators-are-challenging-the-status-quo/.

[10] *See, e.g.*, DEPARTMENT FOR SCIENCE, INNOVATION & TECHNOLOGY, A PRO-INNOVATION APPROACH TO AI REGULATION, 2023, Cm. 815 (UK); INNOVATION, SCI. & ECON. DEV. CAN., THE PAN-CANADIAN AI STRATEGY (2017); SMART NATION SING., NAIS 2.0: SINGAPORE NATIONAL AI STRATEGY (2023).

The strategic significance of AI also resonated within Congress. For instance, the National Artificial Intelligence Initiative Act was proposed in the House of Representatives in December 2020, aiming to secure ongoing U.S. supremacy in AI research and development (R&D).[15] In June 2023, Senators Michael Bennet (D-Colo.), Todd Young (R-Ind.), and Mark Warner (D-Va.) introduced the bipartisan Global Technology Leadership Act.[16] This act was designed to bolster American competitiveness in crucial emerging technologies, including AI. Its need was underscored by the Senators' explicit reference to the technological rivalry with China in these strategic domains.[17]

Reflecting similar concerns, the Chinese government has demonstrated significant ambition in both advancing AI innovation and sculpting its regulatory landscape. As early as 2017, the State Council articulated its aspiration to achieve global leadership in AI by 2030 through its Next-Generation Artificial Intelligence Development Plan.[18] In March 2024, Premier Li

---

[11] Exec. Off. of the President, Nat'l Sci. & Tech. Council Comm. on Tech., Preparing for the Future of Artificial Intelligence (2016), https://obamawhitehouse.archives.gov/sites/default/files/whitehouse_files/microsites/ostp/NSTC/preparing_for_the_future_of_ai.pdf; Nat'l Sci. & Tech. Council Networking & Info. Tech. Rsch. & Dev. Subcomm., National Artificial Intelligence Research and Development Plan (2016), https://www.nitrd.gov/PUBS/national_ai_rd_strategic_plan.pdf.

[12] Exec. Order No. 13859, 84 Fed. Reg. 3967 (Feb. 11, 2019).

[13] Nat'l Sci. & Tech. Council, Select Comm. on A.I., National Artificial Intelligence Research and Development Strategic Plan: 2023 Update (May 2023), https://bidenwhitehouse.archives.gov/wp-content/uploads/2023/05/National-Artificial-Intelligence-Research-and-Development-Strategic-Plan-2023-Update.pdf.

[14] *Id.*

[15] National Artificial Intelligence Initiative Act, H.R. 6216, 116th Cong. (2020).

[16] Global Technology Leadership Act, S. 1873, 118th Cong. (2023).

[17] Michael Bennet, *Bennet, Young, Warner Introduce Bill to Strengthen U.S. Technology Competitiveness*, U.S. Senators for Colo. (June 8, 2023), https://www.bennet.senate.gov/public/index.cfm/2023/6/bennet-young-warner-introduce-bill-to-strengthen-u-s-technology-competitiveness.

[18] Next-Generation Artificial Intelligence Development Plan (新一代人工智能发展规划) (promulgated by the State Council, July 20, 2017, effective July 20, 2017) (China).

Qiang unveiled the "AI Plus" initiative, which aims to harness AI's potential within traditional industries to fuel economic growth and foster technological progress.[19] Additionally, local governments, such as those in Shanghai and Shenzhen, have enacted and implemented regulations to form AI ethics committees. These committees are responsible for overseeing AI development, conducting audits and assessments, and facilitating the creation of industrial parks designed for the easy and lawful exchange of input and training data. Although Chinese legislators are still crafting a comprehensive AI law, a framework of interim regulations is already operational at both the national and local levels.[20]

Despite the exponential growth in AI regulations in both the United States and China, a closer examination of their regulatory frameworks reveals distinct ideological differences between these two global powerhouses. For instance, as depicted in Table 14.1, the AI industry in the United States remains driven by the market and is rather decentralized,[21] while private investors in China tend to depend more on government-guided funds to navigate market trends, a practice influenced by heightened policy uncertainty. On the other hand, some commentators believe that, while the United States views AI competition as a sprint (focusing on exclusivity and short-term victories), China approaches it as a marathon (emphasizing inclusivity and sustained outcomes).[22]

---

[19] Government Work Report of March 5, 2024 at the Second Session of the 14th National People's Congress (政府工作报告—2024年3月5日在第十四届全国人民代表大会第二次会议上), STATE COUNCIL GAZETTE No. 9 [2024]; Ben Jiang, "*Two Sessions" 2024: China's Lawmakers Call for More AI Development to Catch Up with US, While Keeping It Under Regulatory Control*, S. CHINA MORNING POST (Mar. 11, 2024), https://www.scmp.com/tech/policy/article/3254851/two-sessions-2024-china-lawmakers-call-more-ai-development-catch-us-while-keeping-it-under.

[20] *See, e.g.*, Interim Measures for Administration of Generative Artificial Intelligence Services (生成式人工智能服务管理暂行办法) (promulgated by the Cyberspace Admin. of China, Nat'l Dev. & Reform Comm'n, Ministry of Educ., Ministry of Sci. & Tech., Ministry of Indus. & Info. Tech., Ministry of Pub. Sec. & Nat'l Radio & Television Admin., July 10, 2023, effective Aug. 15, 2023) St. Council Gaz., Aug. 30, 2023 (China) [hereinafter Interim Measures for Administration of Generative AI Services]; Provisions on the Administration of Algorithm-Generated Recommendations for Internet Information Services (互联网信息服务算法推荐管理规定) (promulgated by the Cyberspace Admin. of China, Ministry of Indus. & Info. Tech., Ministry of Pub. Sec. & State Admin. for Market Reg., Dec. 31, 2021, effective Mar. 1, 2022) St. Council Gaz., Mar. 30, 2022 (China) [hereinafter Provisions on Algorithm-Generated Recommendations]; Provisions on Administration of Deep Synthesis of Internet-Based Information Services (互联网信息服务深度合成管理规定) (promulgated by the Cyberspace Admin. of China, Ministry of Indus. & Info. Tech. & Ministry of Pub. Sec., Nov. 25, 2022, effective Jan. 10, 2023) St. Council Gaz., Feb. 10, 2023 (China) [hereinafter Provisions on Deep Synthesis].

[21] *See, e.g.*, ANU BRADFORD, DIGITAL EMPIRES: THE GLOBAL BATTLE TO REGULATE TECHNOLOGY (2023); Alex Engler, *The EU and U.S. Diverge on AI Regulation: A Transatlantic Comparison and Steps to Alignment*, BROOKINGS, (Apr. 25, 2023), https://www.brookings.edu/articles/the-eu-and-us-diverge-on-ai-regulation-a-transatlantic-comparison-and-steps-to-alignment/; Emma Klein & Stewart Patrick, *Envisioning a Global Regime Complex to Govern Artificial Intelligence*, CARNEGIE ENDOWMENT FOR INT'L PEACE (Mar. 21, 2024), https://carnegieendowment.org/research/2024/03/envisioning-a-global-regime-complex-to-govern-artificial-intelligence?lang=en; Stewart Patrick, *Rules of Order: Assessing the State of Global Governance*, CARNEGIE ENDOWMENT FOR INT'L PEACE (Sept. 12, 2023), https://carnegieendowment.org/research/2023/09/rules-of-order-assessing-the-state-of-global-governance?lang=en.

**Table 14.1.** Comparison between key performance indicators of the AI industry in the United States and China in 2023

|  | United States | China |
| --- | --- | --- |
| AI patent filings[a] | 32% | *34.7%* |
| Top publications on AI[a] | 22.6% | *36.7%* |
| Number of newly funded AI start-ups[b] | 897 | 122 |
| Private investment in AI[b] | US$67.2 billion | US$7.8 billion |
| Public investment in AI[c] | US$1.8 billion | Not available |
| Notable AI models[b] | 61 | 15 |
| World's top 50 AI hubs[d] | 10 | 2 |

[a] Inst. of Sci. & Tech. Info. of China & Peking Univ., Global AI Innovation Index 2023 [2023全球人工智能创新指数报告] (2023).

[b] Stanford Univ., Inst. for Human-Centered A.I., Artificial Intelligence Index Report 2024 (2024).

[c] Networking & Info. Tech. R&D Prog. & Nat'l A.I. Initiative Off., Supplement to the President's FY 2024 Budget (2023), https://www.nitrd.gov/pubs/FY2024-NITRD-NAIIO-Supplement.pdf.

[d] Bhaskar Chakravorti, Ajay Bhalla, Ravi Shankar Chaturvedi & Christina Filipovic, *50 Global Hubs for Top AI Talent*, Harv. Bus. Rev. (Dec. 21, 2021), https://hbr.org/2021/12/50-global-hubs-for-top-ai-talent.

These differences could shape each country's strengths and weaknesses in the ongoing AI arms race, either enhancing or hindering innovation in AI technologies. More importantly, the escalating technological rivalry has led to exclusionary policymaking, prioritizing national interest over inclusive innovation. With both parties increasingly regulating the flow of technology, capital, and personnel in response to the AI race, ensuring inclusive and sustainable development, as well as equal access to the benefits of AI research, has become more challenging than ever.

In this chapter, we explore three critical aspects of the American and Chinese legal infrastructures that significantly impact AI innovation: data privacy, intellectual property (IP) rights, and export restrictions. Through this comparative analysis, we argue that, while China's legal environment may offer certain advantages in terms of access to training data and IP protection, the United States maintains superior resources by enforcing strict export controls on semiconductor chips, AI models, as well as outbound investments in these areas. This nuanced examination helps illuminate how each country's legal framework could influence the ultimate trajectory of the AI race and how the technological rivalry has led to exclusionary rulemaking on a global scale.

---

[22] Tariq Malik, *The Philosophies Behind the US-China Tech War*, LSE Bus. Rev. (Feb. 5, 2025), https://blogs.lse.ac.uk/businessreview/2025/02/05/the-philosophies-behind-the-us-china-tech-war/.

## Data Privacy

AI technologies and individual privacy intersect primarily in three areas: the training of AI algorithms, automated decision-making (ADM), and government surveillance. Training AI requires very large datasets. It is estimated that computer scientists at OpenAI initially used a 175-billion-parameter model to train ChatGPT-3 in 2019.[23] However, using extensive personal data heightens the risk of privacy breaches.[24] Privacy safeguards in data collection typically hinge on two practices: obtaining informed consent from data subjects and implementing data anonymization by processors. Yet, the widespread adoption of AI could exacerbate privacy risks by increasing the likelihood of data de-anonymization,[25] particularly as larger datasets tend to facilitate this process.[26]

Privacy concerns also emerge during the implementation phase in both ADM and surveillance analytics.[27] For instance, a bank using AI tools in the ADM process to evaluate an applicant's personal data for a loan decision might face issues if the data is incomplete, potentially leading to compromised, biased, or even discriminatory decisions that could severely impact an individual's legal rights or interests.[28] Moreover, integrating AI with surveillance technologies—such as facial recognition systems that identify or verify individuals by their facial features—has intensified privacy intrusions, a longstanding issue in institutional governance.[29] These potential misuses underscore the urgent need for heightened vigilance and proactive measures from AI developers, users, and regulators to safeguard privacy and prevent abuses.

---

[23] Caitlin Chin-Rothmann, *Protecting Data Privacy as a Baseline for Responsible AI*, Ctr. for Strategic & Int'l Stud. (July 18, 2024), https://www.csis.org/analysis/protecting-data-privacy-baseline-responsible-ai.

[24] Shlomit Yanisky-Ravid & Sean K. Hallisey, *"Equality and Privacy by Design": A New Model of Artificial Intelligence Data Transparency via Auditing, Certification, and Safe Harbor Regimes*, 46 Fordham Urb. L.J. 428, 458 (2019).

[25] *Id.* at 471.

[26] *Id.*

[27] Katerina Demetzou, Gabriela Zanfir-Fortuna & Sebastião Barros Vale, *The Thin Red Line: Refocusing Data Protection Law on ADM, A Global Perspective with Lessons from Case-Law*, 49 Comput. L. & Sec. Rev 1, 2 (2023).

## United States

Statutory protections for privacy in the United States vary across sectors and states. Without federal legislation, state laws, sector-specific federal laws, and agency regulations have emerged to address privacy concerns through fragmented and piecemeal solutions.[30] By the end of 2024, twenty states have enacted their own consumer privacy regulations, with the California Privacy Rights Act of 2020 (CPRA), building upon the California Consumer Privacy Act of 2018, being one of the most comprehensive privacy statutes.[31] Introducing the first data minimization rule in the United States, the CPRA emphasized that businesses should only collect consumers' personal information to the extent necessary for the intended purposes. It also mandated an opt-out mechanism for individuals to decline certain forms of ADM systems, including those leveraging AI technologies.[32]

At the federal level, the regulation is fragmented across sectors. Laws such as the Children's Online Privacy Protection Act (COPPA) and the Health Insurance Portability and Accountability Act (HIPAA) hold the potential to regulate certain aspects of AI privacy issues.[33] Yet recent surveys and poll reveal a call from U.S. consumers for a comprehensive federal privacy law in the style of the European Union's General Data Protection Regulation (GDPR).[34] The American Privacy Rights Act was introduced to the U.S. House of Representatives in April 2024, incorporating similar data minimization and opt-out rules.[35] However, the revised draft in June omitted civil rights protection measures that required large data holders, including AI developers, to assess the impact of data usage for model training.[36]

[28] *See, e.g.,* Lee Jyh-An, *Algorithmic Bias and the New Chicago School*, 14 L. Innovation & Tech. 95, 98 (2022); Cameron F. Kerry, *Protecting Privacy in an AI-Driven World*, Brookings (Feb. 10, 2020), https://www.brookings.edu/articles/protecting-privacy-in-an-ai-driven-world/.

[29] Peter Mantello, Manh-Tung Ho, Minh-Hoang Nguyen & Quan-Hoang Vuong, *Machines That Feel: Behavioral Determinants of Attitude Towards Affect Recognition Technology—Upgrading Technology Acceptance Theory with the Mindsponge Model*, 10 Humans. & Soc. Sci. Commc'ns 430 (2023); Wang Xukang, Wu Ying Cheng, Zhou Mengjie & Fu Hongpeng, *Beyond Surveillance: Privacy, Ethics, and Regulations in Face Recognition Technology*, 7 Frontiers Big Data 1, 1 (2024).

[30] Jennifer King & Caroline Meinhardt, *Rethinking Privacy in the AI Era: Policy Provocations for a Data-Centric World* (Stanford Univ., Inst. for Human-Centered A.I., White Paper, Feb. 2024), https://hai.stanford.edu/sites/default/files/2024-02/White-Paper-Rethinking-Privacy-AI-Era.pdf.

[31] Cal. Civ. Code § 1798 (West 2025).

[32] *A New Landmark for Consumer Control Over Their Personal Information: CPPA Proposes Regulatory Framework for Automated Decisionmaking Technology*, Cal. Privacy Prot. Agency (Nov. 27, 2023), https://cppa.ca.gov/announcements/2023/20231127.html.

Despite the absence of a comprehensive federal privacy act, AI developers' data misuse could violate existing antitrust and consumer protection laws, subject to enforcement by the Federal Trade Commission (FTC), if such actions undermine fair competition.[37] Section 5 of the Federal Trade Commission Act prohibits "unfair methods of competition in or affecting commerce,"[38] which, as interpreted by the Supreme Court, extends beyond the conducts specified in the Sherman and Clayton Acts to encompass other unfair practices that negatively affect competition.[39] The FTC has explicitly acknowledged that these unfair methods of competition could include instances where "powerful firms unfairly use AI technologies in a manner that tends to harm competitive conditions."[40] Violations of these regulations may lead to enforcement actions by the FTC. For example, in February 2023, the agency filed a complaint in the Northern District of California against a company that breached its commitment by sharing users' personal information with unauthorized third-party advertising platforms, thereby violating the Federal Trade Commission Act.[41] In July 2023, the FTC imposed a US$7.8 million fine on another company and prohibited it from sharing consumers' health data for advertising purposes.[42]

---

[33] Dennis Crouch, *Using Intellectual Property to Regulate Artificial Intelligence*, 89 Mo. L. Rev. 781, 802 (2024); Karl Manheim & Lyric Kaplan, *Artificial Intelligence: Risks to Privacy and Democracy*, 21 Yale J.L. & Tech. 106, 162 (2019).

[34] Tony Bradley, *Privacy and Trust in the AI Age*, Forbes (Oct. 30, 2024), https://www.forbes.com/sites/tonybradley/2024/10/30/privacy-and-trust-in-the-ai-age/.

[35] American Privacy Rights Act, H.R.8818, 118th Cong. §§ 102, 106 (2024).

[36] *Committee Chairs Cantwell, McMorris Rodgers Unveil Historic Draft Comprehensive Data Privacy Legislation*, U.S. Senate Comm. on Com., Sci. & Transp. (Apr. 7, 2024), https://www.commerce.senate.gov/2024/4/committee-chairs-cantwell-mcmorris-rodgers-unveil-historic-draft-comprehensive-data-privacy-legislation.

[37] Fed. Trade Comm'n, Off. of Tech., *AI Companies: Uphold Your Privacy and Confidentiality Commitments*, Fed. Trade Comm'n (Jan. 9, 2024), https://www.ftc.gov/policy/advocacy-research/tech-at-ftc/2024/01/ai-companies-uphold-your-privacy-confidentiality-commitments.

[38] 15 U.S.C. § 45(a)(1).

[39] *See, e.g.*, Fed. Trade Comm'n v. Ind. Fed'n of Dentists, 476 U.S. 447, 454 (1986); Fed. Trade Comm'n v. Sperry & Hutchinson Co., 405 U.S. 233, 242 (1972); Fed. Trade Comm'n v. Texaco, 393 U.S. 223, 262 (1968); Fed. Trade Comm'n v. Brown Shoe, 384 U.S. 316, 321 (1966); Atlantic Refin. Co. v. Fed. Trade Comm'n, 381 U.S. 357, 369 (1965); Fed. Trade Comm'n v. Colgate-Palmolive Co., 380 U.S. 374, 384–85 (1965); PAN AM v. United States, 371 U.S. 296, 306–08 (1963); Fed. Trade Comm'n v. Nat'l Lead Co., 352 U.S. 419, 428–29 (1957); Am. Airlines, Inc. v. N. Am. Airlines, Inc., 351 U.S. 79, 85 (1956); Fed. Trade Comm'n v. Motion Picture Advert. Serv. Co., 344 U.S. 392, 394–95 (1953); Fed. Trade Comm'n v. Cement Inst., 333 U.S. 683, 708 (1948); Fed. Trade Comm'n v. R.F. Keppel & Bro., Inc., 291 U.S. 304, 310 (1934).

[40] *Comment of the United States Federal Trade Commission*, Fed. Trade Comm'n (Oct. 30, 2023), https://www.ftc.gov/system/files/ftc_gov/pdf/p241200_ftc_comment_to_copyright_office.pdf.

The fragmented, laissez-faire approach to data regulation may inadvertently help maintain the competitiveness of American AI firms, as generalized privacy regulations could stifle digital innovation nationwide.[43] Since AI training inherently involves a trade-off between data availability and privacy safeguards,[44] a more flexible privacy regime ostensibly promises better access to data. Ironically, however, this advantage has gradually been compromised by the overpopulation of such sectoral acts—instead of falling into the gap between regulations, many companies are now facing multiple, overlapping regulations, which makes compliance more complicated.[45]

Meanwhile, there is a growing demand for enhanced privacy safeguards in the AI development process,[46] a factor that could impede the pace of AI innovation in the United States. In October 2022, the Biden administration unveiled the Blueprint for an AI Bill of Rights (Blueprint), encompassing key principles for regulating AI.[47] The Blueprint envisaged that data collection by AI developers should conform to reasonable expectations, to the strictly necessary extent, and with the data subject's consent. It also underscored the importance of safeguarding civil liberties from unwarranted surveillance, advocating for increased oversight of surveillance technologies.[48]

---

[41] Complaint, U.S. v. GoodRx Holdings, Inc., No. 23-cv-460 (N.D. Cal. Feb. 2, 2023).

[42] BetterHelp, Inc., FTC Docket No. C-4796 (Fed. Trade Comm'n July 7, 2023).

[43] Anu Bradford, *The False Choice Between Digital Regulation and Innovation*, 119 Nw. U. L. Rev. 377, 397 (2024).

[44] Ajay Agrawal, Joshua Gans & Avi Goldfarb, Prediction Machines: The Simple Economics of Artificial Intelligence 163 (2018); Peter K. Yu, *Beyond Transparency and Accountability: Three Additional Features Algorithm Designers Should Build into Intelligent Platforms*, 13 Ne. U. L. Rev. 263, 293 (2021).

[45] Daniel J. Solove, *The Growing Problems with the Sectoral Approach to Privacy Law*, TeachPrivacy (Nov. 13, 2015), https://teachprivacy.com/problems-sectoral-approach-privacy-law/.

[46] *See, e.g.*, Tate Ryan-Mosley & Melissa Heikkilä, *Three Things to Know About the White House's Executive Order on AI*, MIT Tech. Rev. (Oct. 30, 2023), https://www.technologyreview.com/2023/10/30/1082678/three-things-to-know-about-the-white-houses-executive-order-on-ai/.

[47] *Blueprint for an AI Bill of Rights Making Automated Systems Work for the American People*, White House (Oct. 2022), https://bidenwhitehouse.archives.gov/ostp/ai-bill-of-rights/.

[48] *Id.*

While the Blueprint advocates for self-regulation within the industry, its nonbinding nature has raised concerns, particularly regarding its exclusion of law enforcement matters. Such exclusion has prompted calls for congressional action to all law enforcement to the proposed regulations.[49] Despite challenges in enforcement, the Blueprint inspired subsequent policymaking and regulatory efforts. In October 2023, the White House issued the Executive Order on the Safe, Secure, and Trustworthy Development and Use of Artificial Intelligence, emphasizing the protection of privacy and civil liberties amidst the rapid advancement of AI technology and highlighting the legal and societal risks posed by improper data collection and use, including the chilling of First Amendment protections.[50] Following these two pivotal initiatives, several federal agencies—such as the FTC, the Department of Defense, the National Institute of Standards and Technology (NIST), the National Telecommunications and Information Administration, the Cybersecurity and Infrastructure Security Agency, and the Securities and Exchange Commission—have integrated AI consideration into their regulatory frameworks.[51]

Following the 2024 Presidential election, the arrival of the Trump administration has profound implications for U.S. AI policy. Notably, Trump repealed Biden administration's executive order, which the new president considered dangerous and potentially stifling to AI innovation.[52] Some commentators anticipate that, with President Trump's re-election, policy priorities will shift toward leveraging AI as a geopolitical advantage, rather than safeguarding individual rights, including privacy.[53] Private stakeholders support this approach. In a Congressional hearing on social media privacy and data abuse, Meta CEO Mark Zuckerberg emphasized the importance of "enabling innovation" when regulating sensitive technologies, such as facial recognition. He specifically referenced threats from Chinese firms that could gain competitive advantages if the extensive regulation hampers the U.S. companies' ability to innovate.[54] Overall, although the lack of an overreaching privacy act appears to favor AI innovation, multiple complications have made this effect less evident.

## China

Since 2021, China has put in place the Personal Information Protection Law (PIPL), the country's first comprehensive personal data protection law.[55] Before the statute's promulgation, privacy protection law in China was as fragmented as in the United States. For example, the Cybersecurity Law of 2016 contains principles of data minimization and informed consent,[56] the E-Commerce Law of 2018 restricts certain ADM practices such as targeted advertising,[57] and the Data Security Law of 2021 mandates certain data processors to employ risk-management measures and conduct periodical risk assessment.[58] In 2020, the National People's Congress passed the new Civil Code, which became effective on January 1, 2021.[59] The Code added privacy and personal information interests to the scope of recognized civil rights, bolstering the enactment and implementation of the PIPL later that year.[60] Multiple guidelines and standards were gradually put in place in the following two years, setting out the specific measures to effectuate the mechanisms pinned down by the PIPL.[61] The legal

---

[54] *Facebook, Social Media Privacy, and the Use and Abuse of Data: Joint Hearing Before the S. Comm. on Com., Sci. & Transp. and the S. Comm. on the Judiciary*, 115th Cong. 22 (2018) (statement of Mark Zuckerberg, Chairman and CEO, Facebook), https://www.congress.gov/event/115th-congress/senate-event/LC64510/text.

[55] Personal Information Protection Law (个人信息保护法) (promulgated by the Standing Comm. Nat'l People's Cong., Aug. 20, 2021, effective Nov. 1, 2021), art. 1 (China).

[56] Cybersecurity Law (网络安全法) (promulgated by the Standing Comm. Nat'l People's Cong., Nov. 7, 2016, effective June 1, 2017), art. 41 (China).

[57] E-Commerce Law (电子商务法) (promulgated by the Standing Comm. Nat'l People's Cong., Aug. 31, 2018, effective Jan. 1, 2019), art. 18 (China).

[58] Data Security Law (数据安全法) (promulgated by the Standing Comm. Nat'l People's Cong., June 10, 2021, effective Sept. 1, 2021), arts. 27, 29, 30 (China).

[59] Civil Code (民法典) (promulgated by the Nat'l People's Cong., May 28, 2020, effective Jan. 1, 2021) (China).

[60] *Id.* arts. 1032, 1034.

[61] *See, e.g.*, Regulations on the Security Protection of Critical Information Infrastructure (关键信息基础设施安全保护条例) (promulgated by the State Council, Apr. 27, 2021, effective Sept. 1,

framework governing personal data protection in China is a comprehensive and vertically integrated regime, with the PIPL at its core. The PIPL grants individuals various rights, including the right to be informed, decide, restrict, or refuse the processing of their personal information, as well as the right to access, copy, transfer, rectify, supplement, or erase their personal data. Moreover, individuals have the right to request that personal information processors clarify their data processing practices.[62] The PIPL also imposes obligations on data processors concerning authorization, notification, consent, security measures, impact assessments, and sharing data with third parties.[63] AI developers and operators are obligated to comply with these legal provisions if they fall under the category of "personal information processors" and "data processors" as defined by the PIPL and the Data Security Law, respectively.[64]

The government's regulation of AI technologies also extends to issues related to data and privacy. In 2022 and early 2023, regulations on certain subcategories of AI technologies such as algorithm-generated recommendations and deep synthesis were released, setting out obligations and liabilities for algorithm operators and deep synthesis service providers.[65] In July 2023, seven state agencies jointly issued the Interim Measures for the Management of Generative Artificial Intelligence Services, the first

2021) St. Council Gaz., Sept. 10, 2021 (China); Measures for the Administration of Data Security in the Field of Industry and Information Technology (for Trial Implementation) (工业和信息化领域数据安全管理办法（试行）), Notice of the Ministry of Indus. & Info. Tech. No. 166 [2022] (promulgated by the Ministry of Indus. & Info. Tech., Dec. 8, 2022, effective Jan. 1, 2023) (China); Opinions on Building Basic Systems for Data to Give Full Play to the Role of Data Resources (关于构建数据基础制度更好发挥数据要素作用的意见) (promulgated by the Cent. Comm'n & State Council, Dec. 2, 2022, effective Dec. 2, 2022) St. Council Gaz., Jan. 10, 2023 (China); Information Security Technology—Guidance for Personal Information Security Impact Assessment (信息安全技术个人信息安全影响评估指南), GB/T 39335-2020 (issued by the Standardization Admin. of China, Nov. 19, 2020, effective June 1, 2021) (China); Information Security Technology—Implementation Guidelines for Notices and Consent in Personal Information Processing (信息安全技术个人信息处理中告知和同意的实施指南), GB/T 42574-2023 (promulgated by the Standardization Admin. of China, May 23, 2023, effective Dec. 1, 2023) (China); Information Security Technology—Personal Information Security Testing and Evaluation Specification in Mobile Internet Applications (App) (信息安全技术移动互联网应用程序（App）个人信息安全测评规范), GB/T 42582-2023 (issued by the Standardization Admin. of China, May 23, 2023, effective Dec. 1, 2023) (China).

[62] Personal Information Protection Law, *supra* note 55, arts. 44–48.
[63] *Id.* arts. 13, 14, 17, 51–57.
[64] Shen Weixing & Liu Yun, *China's Normative Systems for Responsible AI: From Soft Law to Hard Law, in* The Cambridge Handbook of Responsible Artificial Intelligence: Interdisciplinary Perspectives 150, 157 (2022).
[65] Provisions on Algorithm-Generated Recommendations, *supra* note 20, art. 7; Provisions on Deep Synthesis, *supra* note 20, art. 14.

comprehensive AI regulation in China.[66] The Interim Measures contain rules safeguarding AI users' rights to their personal information, as well as requirements on data minimization, anonymization, and prohibition against unauthorized sharing.[67] Furthermore, it sets forth requirements regarding training data of machine learning models, including those requiring these training data to have a legitimate source with respect to IP and personal information.[68]

Under the general rule in the PIPL, state agencies are held to the same data protection standards as private data processors. However, the statute provides certain exceptions—for instance, the waiver of the notification requirement if providing notice would impede the state agency's ability to fulfill its legal obligations.[69] These exceptional provisions may permit the deployment of surveillance technologies, including image capture and personal identification systems, in public spaces to enhance public security,[70] even though biometric data is classified as "sensitive personal information" and requires a higher level of protection.[71] As a result, the notion of "public security" is often perceived as a manipulable tool, as governments could broadly interpret this concept to bypass necessary procedures.[72] Consequently, many of the envisioned mechanisms for safeguarding personal data outlined in laws and regulations may not function as effectively in practice.[73] Some critics go as far as to characterize privacy legislation in China as granting the state a near-exclusive authority over the permissible use of personal information, rather than primarily focusing on the protection of individuals' data privacy.[74]

Given that government-collected data typically surpasses private-sector data in both volume and breadth, AI companies that can tap into government-controlled data gain a significant edge in training their language models.[75] A key avenue for private firms to secure such data access is by offering services to governmental entities.[76] In China, prominent technology firms foster cooperative ties with government bodies by engaging in government-driven initiatives, while regulatory oversight and standardization mechanisms empower the government to steer the industry.

---

[66] Interim Measures for Administration of Generative AI Services, *supra* note 20.

[67] *Id.* art. 11.

[68] *Id.* art. 7.

[69] Personal Information Protection Law, *supra* note 55, art. 35.

[70] *Id.* art. 26.

[71] *Id.* art. 28.

[72] Lee Jyh-An & Zhou Peng, *FRT Regulation in China*, *in* The Cambridge Handbook of Facial Recognition in the Modern State 242, 246 (Rita Matulionyte ed., 2024).

[73] *Id.* at 247.

[74] Karman Lucero, *Artificial Intelligence Regulation and China's Future*, 33 Colum. J. Asian L. 94, 142 (2019).

[75] Martin Beraja, David Y. Yang & Noam Yuchtman, *Data-Intensive Innovation and the State: Evidence from AI Firms in China*, 90 Rev. Econ. Stud. 1701, 1701 (2023).

Collaboration between public and private sectors manifests across various AI applications, spanning from crime detection to hotel check-ins to hospital registrations. Facial recognition stands out as a prime example among these applications, as the PIPL authorizes Chinese public security agencies to capture facial data through surveillance systems. Subsequently, by partnering with public security units, AI companies can gain access to government surveillance data, leveraging them to enhance and innovate their algorithms.[77]

Empirical evidence underscores the significant impact of government data on commercial AI advancement, highlighting how furnishing government data to AI companies serving state interests has propelled Chinese entities to the forefront of global facial recognition technology.[78] China hosts leading facial recognition enterprises like Megvii and SenseTime, based in Beijing and Hong Kong, respectively. The existence of these technology firms and their success underscore the potential of readily available government data as an innovation catalyst that can drive AI progress within a nation.[79] In contrast, major U.S. technology firms, such as Microsoft, Amazon, and IBM, have refrained from supplying facial recognition solutions to U.S. law enforcement following instances of police misconduct that sparked widespread protests.[80] Such actions may impede the evolution of AI technologies in these spheres, signaling a divergence in approaches between the United States and China regarding the utilization of facial recognition technologies.

## Copyright

The intersection of AI and copyright primarily revolves around two aspects: (1) at the input stage, whether the use of copyright-protected content in AI training constitutes copyright infringement; and (2) at the output stage, whether AI-generated content is copyrightable.

---

[76] *Id.* at 1702.

[77] *Id.*

[78] *Id.* at 1720.

[79] *Id.* at 1721.

[80] Rebecca Heilweil, *Big Tech Companies Back Away from Selling Facial Recognition to Police. That's Progress*, Vox (June 12, 2020), https://www.vox.com/recode/2020/6/10/21287194/amazon-microsoft-ibm-facial-recognition-moratorium-police; Karen Weise & Natasha Singer, *Amazon Pauses Police Use of Its Facial Recognition Software*, N.Y. Times (June 10, 2020), https://www.nytimes.com/2020/06/10/technology/amazon-facial-recognition-backlash.html.

## Using Copyrighted Work as Training Data

Potential copyright infringement may arise during the training phase,[81] with AI developers ordinarily defending their use of other's copyrighted works based on limitations or exceptions in copyright law. If AI developers and users are forced to abandon their projects due to the threat of wide-spread copyright infringement, this uncertainty could adversely impact the growth of the AI industry. Consequently, the extent to which AI developers face copyright infringement risks will significantly influence the pace of AI development.

In the United States, fair use, the primary form of copyright limitation and exception, requires the determination of four key factors: the purpose and character of the use, the nature of the copyrighted work, the amount and substantiality of the portion used in relation to the whole work, and the effect of the use on the potential market for or value of the copyrighted work.[82] The introduction of the transformative use under the first factor has been pivotal in many fair use cases.[83]

A fair use analysis can yield conclusions on both sides of the debate. Some scholars argue that AI models do not undergo significant enough transformation to qualify as fair use, as many of these models do not produce a creative output that is distinctly different in nature or purpose from the original work.[84] Additionally, the third and fourth factors can present obstacles for AI training to qualify as fair use. Copyright owners may contend that courts should rule in favor of the third factor, as machine learning often involves replicating the entire work.[85] Moreover, given the possibility for copyrighted works used for AI training to represent a lucrative licensing market, which enables copyright owners to generate significant revenues, AI developers

---

[81] *See, e.g.*, Christina Han, *Parasites or Boosters of Human Creativity?: A Model to Solve Copyright Issues of Training Dataset of Artificial Intelligence Before AlphaArt Defeats Human Artists*, 49 AIPLA Q.J. 281, 296 (2021) (stating that extracting training dataset from scrapped content from the Internet could potentially amount to copyright infringement if the content is copied without authorization).

[82] 17 U.S.C. § 107.

[83] Pierre N. Leval, *Toward a Fair Use Standard*, 103 HARV. L. REV. 1105, 1111 (1990).

[84] Mark A. Lemley & Bryan Casey, *Fair Learning*, 99 TEX. L. REV. 743, 764 (2021).

[85] *Id*. at 763–65.

may find it challenging to make a compelling argument concerning the fourth factor.[86]

Conversely, others argue that using copyrighted materials to train generative AI models should be considered fair use due to its transformative nature.[87] Furthermore, although the works in the training data are usually copied in their entirety, such complete copying is merely an intermediate technical step in the analytical process, unrelated to communicating the underlying original expression. As long as this use remains nonexpressive, it will not have a cognizable effect on the potential market for or value of the copyrighted work.[88]

While arguments can be made for both sides, it is indisputable that determining fair use under U.S. law is a highly fact-specific exercise. This inherent legal uncertainty can induce substantial infringement risks, potentially discouraging AI development within the country. In *Andersen v. Stability AI Ltd.*, a group of artists filed a class action lawsuit against several AI companies, including Stability AI, the developer of AI image generator Stable Diffusion. They alleged, inter alia, copyright infringement for the AI's generation of works substantially similar to the artists' composition style.[89] In response to the defendants' motion to dismiss, the court allowed the plaintiffs' copyright infringement claim to go forward even though it dismissed several claims based on breach of contract, unjust enrichment, and the Digital Millennium Copyright Act.[90] Denying the defendants' motion, the court ruled that "whether a fair use defense applies are issues that must be tested on an evidentiary basis."[91]

While all cases regarding the unauthorized utilization of copyrighted works as training data are currently pending in the United States, it is evident that the ability of AI companies to effectively defend such use as fair use will have a crucial impact on their operational expenses and the pace of technological advancement in the AI field moving forward.

---

[86] *Id.* at 765–66.
[87] *See, e.g.*, Elizabeth Spica, *Public Interest, the True Soul: Copyright's Fair Use Doctrine and the Use of Copyrighted Works to Train Generative AI Tools*, 33 TEX. INTELL. PROP. L.J. 67, 80–87 (2024).
[88] Matthew Sag & Peter K. Yu, *The Globalization of Copyright Exceptions for AI Training*, 74 EMORY L.J. 1163 (2025).
[89] Andersen v. Stability AI Ltd., No. 23-cv-00201-WHO, 2024 WL 3823234, at *1 (N.D. Cal. Aug. 12, 2024).
[90] *Id.*
[91] *Id.*

The Chinese Copyright Law specifies thirteen distinct limitations and exceptions.[92] To be deemed noninfringing in China, a use must fall into one of the thirteen categories enumerated in Article 24 of the Copyright Law—such as teaching, research, or news reporting. At first glance, this enumerative approach may seemingly disadvantage AI developers seeking to invoke fair use to mitigate liability, as AI does not fall squarely within the specified categories of Article 24. While the exceptions for "private research" and "education and public-school research" might encompass certain uses of training data, both exceptions explicitly exclude commercial exploitation and impose limitations on the scale of use, which AI training is likely to surpass.[93]

Nevertheless, Article 24 includes at the end a catch-all provision, stipulating that "other circumstances provided in laws and administrative regulations" may be exempted from copyright infringement liabilities.[94] This miscellaneous provision has partially shifted the closed-list approach toward an open-ended one in copyright law by providing more flexibility for copyright authorities to decide new copyright limitations and exceptions via administrative rules.[95] Therefore, the copyright authorities may play a pivotal role in fostering the advancement of AI through the expansion of copyright limitations and exceptions.

Such flexibility is not currently the reality in China. In 2023, the Guangzhou Internet Court found an AI company liable for generating images that were substantially similar to the Japanese anime figure Ultraman, owned by Tsuburaya Productions.[96] There was a high likelihood that the AI company used Tsuburaya's Ultraman characters to train its AI model. The case revealed the significant legal risks associated with copyright infringement for AI companies operating in China.

Presently, both U.S. and Chinese AI companies face the risk of copyright infringement arising from the unauthorized use of copyrighted works as training data. However, further clarifications from U.S. and Chinese courts and copyright authorities could mitigate this risk. In essence, these decisions will influence the AI race between the two nations.

---

[92] Copyright Law (著作权法) (promulgated by the Standing Comm. Nat'l People's Cong., Sept. 7, 1990, rev'd Nov. 11, 2020, effective Nov. 11, 2020), art. 24 (China).

[93] He Tianxiang, *Copyright Exceptions Reform and AI Data Analysis in China: A Modest Proposal*, in Artificial Intelligence and Intellectual Property 196, 205 (Lee Jyh-An, Reto Hilty & Liu Kung-Chung eds., 2021) [hereinafter AI and Intellectual Property].

[94] Copyright Law, *supra* note 92, art. 24(13).

[95] Lee Jyh-An & Li Yangzi, *The Pathway Towards Digital Superpower: Copyright Reform in China*, 70 GRUR Int'l 861, 867 (2021).

[96] Shanghai Xinchuanghua Cultural Dev. Co. v. Guangzhou XX Network Tech. Co. (上海新创华文化发展有限公司诉广州XX网络科技有限公司著作权侵权纠纷案), (2024) Yue 0192 Min Chu No. 113 ((2024)粤0192民初113号) (Guangzhou Internet Ct. Feb. 8, 2024) (China).

## Copyrightability of AI-Generated Content

The eligibility of AI-generated works for copyright protection significantly impacts the incentives for AI development.[97] Currently, in most jurisdictions, AI-generated works are not subject to copyright protection due to their lack of human creativity, which is essential for originality—a fundamental requirement for copyright protection.[98] This principle was affirmed in the United States when the U.S. District Court for the District of Columbia—and, later, also the U.S. Court of Appeals for the District of Columbia Circuit—ruled that a work of art created solely by AI, without any human input, could not be copyrighted.[99] This decision upheld the Copyright Office's rejection of an application filed by computer scientist Stephen Thaler for an artwork entitled "A Recent Entrance to Paradise" created by a computer algorithm. In his application, Thaler identified the authorship as "autonomously created by a computer algorithm running on a machine" and claimed copyright himself under the work-made-for-hire doctrine, given his ownership of the computer algorithm.[100]

In rejecting Thaler's claim, the district court reasoned that, under the 1976 Copyright Act, copyright subsists in "original works of authorship fixed in any tangible medium of expression, now known or later developed, from which they can be perceived, reproduced, or otherwise communicated, either directly or with the aid of a machine or device."[101] The statute requires that the work be fixed "by or under the authority of the author."[102] The court pointed out that while the Copyright Act does not explicitly define the term "author," the term's linguistic meaning denotes an originator with the capacity for intellectual, creative, or artistic labor. This originator must be a human being, as "human authorship is a bedrock requirement of copyright."[103] Thus, the court rejected Thaler's claim for copyright.[104]

---

[97] *See, e.g.,* Jeremy A. Cubert & Richard G.A. Bone, *The Law of Intellectual Property Created by Artificial Intelligence, in* RESEARCH HANDBOOK ON THE LAW OF ARTIFICIAL INTELLIGENCE 411, 425 (Woodrow Barfield & Ugo Pagallo eds., 1st ed. 2018); Jane C. Ginsburg & Luke A. Budiardjo, *Authors and Machines*, 34 BERKELEY TECH. L.J. 343 (2019); Shlomit Yanisky-Ravid & Luis Antonio Velez-Hernandez, *Copyrightability of Artworks Produced by Creative Robots, Driven by Artificial Intelligence Systems and the Originality Requirement: The Formality-Objective Model*, 19 MINN. J.L. SCI. & TECH. 1 (2018).

[98] *See, e.g.,* Lee Jyh-An, *Computer-Generated Works Under the CDPA 1988, in* AI AND INTELLECTUAL PROPERTY, *supra* note 93, at 177, 183; Jani Ihalainen, *Computer Creativity: Artificial Intelligence and Copyright*, 13 J. INTELL. PROP. L. & PRAC. 724, 726–27 (2018); Paul Lambert, *Computer-Generated Works and Copyright: Selfies, Traps, Robots, AI and Machine Learning* 39 EUR. INTELL. PROP. REV. 12, 14 (2017); Mark Perry & Thomas Marhoni, *From Music Tracks to Google Maps: Who Owns Computer-Generated Works?* 26 COMPUT. L. SEC. REV 621, 624–25 (2010).

[99] Thaler v. Perlmutter, 687 F. Supp. 3d 140 (D.D.C. Aug. 18, 2023), *aff'd*, 130 F.4th 1039 (D.C. Cir. 2025).

[100] *Id.* at 143.

Notably, the United States Copyright Office has previously denied applications from AI users for creations generated with AI tools.[105] The Office explained that creators using AI text-to-image generators cannot meet the human authorship requirement if the work was produced through a process involving random noise, where the creator did not dictate the specific results.[106] This stance was further affirmed in its "Copyright and Artificial Intelligence" report, which concluded that even an extremely detailed or complex prompt does not confer copyright ownership over an AI-generated output.[107]

Nevertheless, human users can still secure copyright over a compilation of AI-generated works if they infuse it with originality. In February 2025, the United States Copyright Office for the first time granted copyright to an AI-generated image, "A Single Piece of American Cheese."[108] Initially, the copyright registration was rejected, but the decision was subsequently reversed. The Office recognized that the work "contains a sufficient amount of human original authorship in the selection, arrangement, and coordination of the AI-generated material that may be regarded as copyrightable."[109] Notably, although the initial image was produced by an AI program named Invoke, the user added original expression by repeatedly selecting, modifying, and regenerating parts of the image. This process transformed the initial prompt-based AI-generated outputs into an "original work guided by human creativity," and the user has recorded every step of the transformation.[110] Following this landmark registration, Assistant General Counsel Jalyce Mangum revealed in an April 2025 interview that the Office "has registered more than a thousand works where applicants have followed our guidance to disclose and disclaim AI-generated material."[111] Therefore, this decision from the Copyright Office sends a positive signal to artists across the United States who incorporate AI into their creative processes.

---

[101] 17 U.S.C. § 102(a).

[102] *Thaler*, 687 F. Supp. 3d at 147.

[103] *Id*. at 146.

[104] *Id*. at 150.

[105] *See, e.g.*, Letter from the Copyright Rev. Bd. to Tamara Pester, Esq., Tamara S. Pester, LLC (Sept. 5, 2023), https://www.copyright.gov/rulings-filings/review-board/docs/Theatre-Dopera-Spatial.pdf [hereinafter Copyright Review Board Letter to Tamara Pester]; Letter from the U.S. Copyright Off. to Van Lindberg, Taylor English Duma LLP (Feb. 21, 2023), https://www.copyright.gov/docs/zarya-of-the-dawn.pdf [hereinafter USCO Letter to Van Lindberg].

[106] Edward Lee, *Prompting Progress: Authorship in the Age of AI*, 76 Fla. L. Rev. 1445, 1454 (2024).

[107] U.S. Copyright Off., Copyright and Artificial Intelligence: Part 2: Copyrightability 18–20 (2025).

[108] Kent Keirsey, *Invoke Secures Copyright in Landmark Ruling for AI-Assisted Artwork*, Invoke Blog (Feb. 10, 2025), https://www.invoke.com/post/invoke-receives-copyright-in-landmark-ruling-for-ai-assisted-artwork.

[109] *How We Received the First Copyright for a Single Image Created Entirely with AI-Generated Material*, Invoke Blog (Feb. 10, 2025), https://44037860.fs1.hubspotusercontent-na1.net/hubfs/44037860/Invoke-First-Copyright-Image-AI-Generated-Material-Report.pdf.

In contrast to the United States, China has adopted a more lenient approach by recognizing the copyrightability of AI-generated works. In *Shenzhen Tencent Computer System Co. v. Shanghai Yingxun Technology Co.*, Tencent successfully argued that human originality was present in the AI-generated work, thereby making it eligible for copyright protection.[112] This case involved a commentary on the stock market composed by AI software developed by Tencent. Unlike the view taken by the *Thaler* court in the United States, the Chinese court in *Tencent* held that human originality could be identified in various phases of the AI software's process of creating the article.[113] The court ruled that, although the AI system took only two minutes to produce the disputed article without any human participation, its automatic operation did not occur in a vacuum. The software was not self-aware; rather, its autonomous operation reflected the developers' personalized selection and arrangement of data types, data formats, the conditions triggering the writing of the article, the templates for article structure, the setting of the corpus, and the training of the intelligent verification algorithm model.[114]

---

[110] *Id.*

[111] Miriam Lord, *US Copyright Office on AI: Human Creativity Still Matters, Legally*, WIPO Mag. (Apr. 24, 2025), https://www.wipo.int/web/wipo-magazine/articles/us-copyright-office-on-ai-human-creativity-still-matters-legally-73696.

[112] Shenzhen Tencent Comput. Sys. Co. v. Shanghai Yingxun Tech. Co. (深圳腾讯诉上海盈讯著作权侵权案), (2019) Yue 0305 Min Chu No. 14010 ((2019)粤0305民初14010号) (Shenzhen Nanshan Dist. Ct. Dec. 24, 2019) (China).

[113] *Id.*

[114] *Id.*; *see also* Rita Matulionyte & Lee Jyh-An, *Copyright in AI-Generated Works: Lessons from Recent Developments in Patent Law*, 19 SCRIPTed 5, 16–18 (2022) (providing more details of the case).

In another case, *Li v. Liu*, the plaintiff created an image by entering specific prompts into the AI system Stable Diffusion and subsequently shared this image on his social media platform.[115] It later came to the plaintiff's attention that the defendant had utilized this image in her own article on a different social media site without obtaining permission. In assessing the originality of the contested image, the Beijing Internet Court aligned its reasoning with that of the *Tencent* court. The Beijing court noted that "the plaintiff employed prompt words to influence elements of the image such as the character portrayal and its presentation, and adjusted parameters to refine the layout and composition of the image . . . Such modifications and personalization indicate the plaintiff's aesthetic sensibility and individual judgment."[116] Accordingly, the court recognized the originality in the AI-generated image and the plaintiff as the copyright owner.[117]

However, in another case, *Feng v. Zhu*, the Suzhou Intermediate People's Court rejected the plaintiff's copyright claim over images generated by Midjourney, a text-to-image AI application, on the grounds that she failed to provide sufficient evidence of her own creative input during the creation process.[118] The plaintiff herself conceded that it was impossible to reproduce the exact same AI-generated images, even when using identical prompts, due to the inherent randomness in Midjourney's generative process. The court further elaborated that works primarily generated automatically by AI are unlikely to qualify for copyright protection, as it is AI—not the user—that ultimately determines the output. Conversely, if the user employs AI merely as a tool and exercises original intellectual input throughout the creative process, the resulting AI-assisted content may be copyrightable.

The perspectives on originality in AI-generated works vary significantly between the United States and China, a disparity that could potentially influence the competitive dynamics of AI development between these two nations. In this context, several American scholars have criticized the stance taken by the United States Copyright Office and federal courts in relation to the copyright registration applications in cases such as those involving Stephen Thaler,[119] *Zarya of the Dawn*,[120] Jason Allen,[121] and *Suryast*.[122] They argue that these decisions hinder the advancement of AI technologies, which enable new and more accessible modes of creative production.[123] These critics contend that the stringent criteria for AI originality in the United States may place domestic creators and businesses at a considerable competitive disadvantage.[124] Conversely, Chinese courts have taken proactive steps to encourage AI innovation by recognizing copyright in AI-generated works. Despite the United States Copyright Office's acknowledgment of the originality of human users in the selection, coordination, and arrangement of AI-generated works, the AI-generated content itself remains uncopyrightable under the U.S. perspective. Therefore, it can be reasonably argued that China has created a more conducive environment for the copyright protection of AI-generated works.

---

[115] Li v. Liu (李某某诉刘某某侵害作品署名权、信息网络传播权纠纷案), (2023) Jing 0491 Min Chu No. 11279 ((2023)京0491民初11279号) (Beijing Internet Ct. Nov. 27, 2023) (China).
[116] *Id.*
[117] *Id.*
[118] Feng v. Zhu (丰某诉朱某等著作权侵权、不正当竞争纠纷案), (2025) Su 05 Min Zhong No. 4840 ((2025)苏05民终4840号) (Suzhou Interm. People's Ct., Apr. 17, 2025) (China) .

## Export Restrictions

As the AI race intensifies, a country's policy aims not only to promote the development of its own AI sector but also to hinder that of its competitors. This regulatory approach extends beyond AI algorithms and models to encompass the entire upstream and downstream infrastructure. This strategy is evident in the countries' efforts to dominate control over capital, knowledge, resources, raw materials, and personnel in the relevant fields.

One of the primary battlegrounds between the United States and China centers on semiconductor chips, which are crucial to the AI industry. An AI chip consists of graphics processing units (GPUs), field-programmable gate arrays (FPGAs), and application-specific integrated circuits (ASICs) specialized for AI. As AI advances, general-purpose chips such as central processing units (CPUs) are becoming less favorable, since AI chips are tens or thousands of times faster, more efficient, and more cost-effective than CPUs for training AI algorithms.[125] Furthermore, older AI chips with larger, slower, and more energy-consuming transistors incur more costs, which will quickly rise to unaffordable levels, making it virtually impossible to develop cutting-edge AI algorithms without highly advanced AI chips.[126] These challenges led to countries' determination to develop the most advanced AI chips amid the AI war.

---

[119] *Thaler v. Perlmutter*, 687 F. Supp. 3d 140 (D.D.C. Aug. 18, 2023), *aff'd*, 130 F.4th 1039 (D.C. Cir. 2025).

[120] USCO Letter to Van Lindberg, *supra* note 105.

[121] Copyright Review Board Letter to Tamara Pester, *supra* note 105.

[122] Letter from the Copyright Rev. Bd. to Alex P. Garens, Esq., Day Pitney, LLP (Dec. 11, 2023), https://www.copyright.gov/rulings-filings/review-board/docs/SURYAST.pdf.

[123] Lee, *supra* note 106, at 1517.

[124] *Id*. at 1563–64.

Notably, both the United States and China have provided substantial fiscal support to the development of advanced AI and semiconductor chips. According to the National Science and Technology Council's report, U.S. government agencies allocated US$1.8 billion to AI R&D in the fiscal year 2023, with a larger budget of US$1.9 billion requested for the fiscal year 2024.[127] In addition to direct funding for AI firms, subsidies are also provided to upstream manufacturers of semiconductor chips to "bring cutting-edge artificial-intelligence chip development and manufacturing to American soil."[128] This funding was allocated under the CHIPS and Science Act of 2022, which requires recipients to "demonstrate significant worker and community investments" and prohibits them from "build[ing] certain facilities in China and other countries of concern."[129] Recipients of subsidies under this act include domestic chip makers such as Intel and foreign manufacturers like Taiwan Semiconductor Manufacturing Co. (TSMC) and Samsung as an effort to entice them to build factories and bring the most advanced technology to the United States.[130]

In addition, the government has concurrently implemented measures to restrict access to these critical resources for China. Since 2018, the

U.S. government has employed export controls that impact the AI sector, affecting everything from semiconductor chip production to AI modeling and financial investment. In 2022, the Department of Commerce's Bureau of Industry and Security (BIS) introduced export controls on specific advanced computing chips and semiconductor chip manufacturing equipment, as well as transactions involving supercomputers and certain integrated circuit applications.[131] These restrictions were further expanded in 2023 to include less advanced chips previously not covered by the bans.[132]

This strategic move aligns with the U.S. response to China's ambitious goal to become a global AI leader by 2030. The BIS articulated that these restrictions are designed to safeguard U.S. national security and uphold foreign policy interests, while also emphasizing that U.S. technological supremacy is rooted in both values and innovation.[133] Under these regulations, advanced chips such as Nvidia's A100 and H100 data center chips and their subsequent models tailored for Chinese markets post-2022 ban, the A800 and H800 GPUs, are prohibited from being sold to China.[134] Reports indicate that the Department of Commerce has also mandated foreign firms like TSMC to cease supplying China with seven-nanometer or more advanced chips.[135] These measures are poised to significantly impact major Chinese technology firms, including Tencent, ByteDance, and Alibaba.[136]

Officials have expressed that the objective of these export limitations is to curtail China's access to advanced semiconductor chips, which are pivotal to AI advancements, and sophisticated computing systems crucial for military applications in China.[137] The regulations were further intensified in April and December 2024, with the December update incorporating twenty-four types of semiconductor chip manufacturing equipment and three types of software tools essential for the development or production of semiconductor chips, thereby bolstering the efficacy of existing controls.[138] Furthermore, international allies such as Japan and the Netherlands have joined the United States in imposing similar controls on semiconductor chip equipment, effectively preventing China from acquiring chip-making technology from alternative sources, including industry giants like ASML, Nikon, and Tokyo Electron.[139]

---

[131] *Commerce Implements New Export Controls on Advanced Computing and Semiconductor Manufacturing Items to the People's Republic of China* (PRC), Bureau of Indus. & Sec. (Oct. 7, 2022), https://www.bis.doc.gov/index.php/documents/about-bis/newsroom/press-releases/3158-2022-10-07-bis-press-release-advanced-computing-and-semiconductor-manufacturing-controls-final/file [hereinafter *Commerce Implements New Export Controls*].

[132] *Implementation of Additional Export Controls: Certain Advanced Computing Items; Supercomputer and Semiconductor End Use; Updates and Corrections*, Bureau of Indus. & Sec. (Oct. 16, 2023), https://www.bis.doc.gov/index.php/documents/federal-register-notices-1/3353-2023-10-16-advanced-computing-supercomputing-ifr/file.

[133] *Commerce Implements New Export Controls*, *supra* note 131.

[134] Che Pan & Ann Cao, *How the Latest US Chip Export Controls Exposed China's Weak Link in the Semiconductor Supply Chain*, S. China Morning Post (Nov. 4, 2023), https://www.scmp.com/tech/tech-war/article/3239803/how-latest-us-chip-export-controls-exposed-chinas-weak-link-semiconductor-supply-chain.

[135] Karen Freifeld & Fanny Potkin, *Exclusive: US Ordered TSMC to Halt Shipments to China of Chips Used in AI Applications*, Reuters (Nov. 10, 2024), https://www.reuters.com/technology/us-ordered-tsmc-halt-shipments-china-chips-used-ai-applications-source-says-2024-11-10/.

[136] Pan & Cao, *supra* note 134.

In the future, the scope of export control will likely be broadened to include downstream AI models.[140] The proposed Enhancing National Frameworks for Overseas Restriction of Critical Exports (ENFORCE) Act, pending approval by both the House of Representatives and the Senate, seeks to amend the Export Control Act of 2018. This amendment would empower the BIS and the White House to require U.S. companies to secure licenses for exporting AI models to China that could threaten U.S. national security.[141]

The U.S. government has also imposed restrictions on domestic companies' investments in foreign chip and AI industries. In August 2023, the Biden administration issued an executive order prohibiting U.S. investments in China's advanced chip and AI sectors.[142] The White House explained that, while the United States generally supports international investment, it must align with national security interests. The administration expressed concerns that certain investments could potentially enhance the development of sensitive technologies in nations that might use them to undermine U.S. and allied capabilities.[143] Consequently, foreign investment in China's semiconductor sector plummeted to US$600 million in 2023, marking its lowest level since 2020.[144]

In response to the escalating export controls imposed by the United States, particularly concerning semiconductor chips, China has implemented retaliatory measures that include restricting exports of essential raw materials for semiconductor chip production and substantially financing its domestic chipmaker. In July 2023, the Chinese Ministry of Commerce announced export controls on critical rare earth materials such as gallium and germanium.[145] In October 2023, China tightened existing export restrictions on graphite by applying more stringent standards.[146] A similar ban on antimony and superhard materials was introduced in August 2024.[147] Four months later, China further strengthened these restrictions, just a day after the United States announced its latest comprehensive export controls.[148] Unlike previous measures, which the Ministry stated did not target any specific country or region,[149] this latest action was explicitly retaliatory against the United States. According to the U.S. Geological Survey, China accounts for 98 percent of the world's gallium production and supplied more than half of the U.S. germanium imports from 2019 to 2022.[150] These raw materials are essential to the manufacture of high-frequency computer chips, electronic vehicle batteries, and national defense systems.

---

[143] *Id.*

[144] The market decline occurred even before the policy's promulgation in August, right after its announcement. Che Chang & John Liu, *"De-Americanize": How China Is Remaking Its Chip Business*, N.Y. Times (May 11, 2023), https://www.nytimes.com/2023/05/11/technology/china-us-chip-controls.html.

[145] Notice on Implementing Export Control on Gallium- and Germanium-Related Substance (关于对镓、锗相关物项实施出口管制的公告), Notice of the Gen. Admin. of Customs No. 3 [2023] (issued by the General Admin. of Customs of the Ministry of Com., July 3, 2023, effective Aug. 1, 2023) (China).

[146] Notice on Improving and Adjusting the Provisional Export Control Measures on Graphite (关于优化调整石墨物项临时出口管制措施的公告), Notice of the Gen. Admin. of Customs No. 39 [2023] (promulgated by the Gen. Admin. of Customs of the Ministry of Com., Oct. 20, 2023, effective Dec. 1, 2023) (China).

[147] Notice on Implementing Export Control on Antimony and Other Related Substance (关于对锑等相关物项实施出口管制的公告), Notice of the Gen. Admin. of Customs No. 33 [2024] (promulgated by the Gen. Admin. of Customs of the Ministry of Com., Aug. 15, 2024, effective Sept. 15, 2024) (China).

[148] Notice on Strengthening Export Control of Dual-Use Substance to the United States (关于加强相关两用物质对美国出口管制的公告), Notice of the Gen. Admin. of Customs No. 46 [2024] (promulgated by the Gen, Admin. of Customs of the Ministry of Com., Dec. 3, 2024, effective Dec. 3, 2024) (China).

[149] *Press Conference of the Spokesperson of the Ministry of Commerce on the Export Control Policy for Antimony and Other Related Substance* (商务部新闻发言人就锑等物项出口管制政策应询答记者问), Ministry of Com. (Aug. 15, 2024).

[150] U.S. Dep't of the Interior, U.S. Geological Surv., Mineral Commodities Summary 2024, at 75, 80 (2024), https://pubs.usgs.gov/periodicals/mcs2024/mcs2024.pdf.

In response to China's export control, the U.S. Department of Defense acknowledges that, while there may be short-term consequences, the government is actively enhancing domestic mining operations and diversifying supply chains.[151] Potential alternative sources include resource-rich nations such as Australia and Canada, although the cost of materials from these countries is anticipated to rise.[152] Despite these challenges, in recognition of the risks associated with excessive dependence on foreign critical minerals, President Trump signed an executive order in April 2025 amid the ongoing tariff war, directing the Secretary of Commerce to initiate a Section 232 investigation and make relevant recommendations. Industry observers suggest that these export restrictions imposed by China may ultimately encourage U.S. lawmakers to augment investments in critical minerals.[153]

In addition to implementing retaliatory export controls, the Chinese government has been actively cultivating its domestic chipmaking industry. In 2014, the State Council issued the Guidelines on Promoting National Integrated Circuit Development, leading to the creation of the China Integrated Circuit Industry Investment Fund, commonly known as "the Big Fund." This initiative aimed to bolster integrated circuit manufacturing within the nation, drawing initial investments from several of the largest state-owned enterprises in the communications and finance sectors. Notable beneficiaries have included leading domestic chipmakers such as Semiconductor Manufacturing International Corporation (SMIC), a state-owned entity, and Hua Hong, a key supplier to Huawei. In May 2024, the Big Fund entered its third phase, amassing a registered capital of RMB 344 billion (approximately US$47.09 billion), with all six major commercial banks participating as investors.[154] This state-backed fund has compensated

for the withdrawal of foreign capital. In 2023, China surpassed the United States in investment into semiconductor startups by a significant margin.[155] Highlighting this progress, Huawei, in August 2023, introduced its Mate 60 Pro, which features a seven-nanometer processor from SMIC, developed amidst stringent U.S. export controls. With this fresh influx of capital, SMIC is now fervently advancing its R&D toward more sophisticated five-nanometer chips, positioning itself just one generation behind the forefront of three-nanometer technology.[156]

Commentators have pointed out that China's substantial investment in semiconductor R&D may place the United States in a challenging predicament. While easing export controls could grant China access to state-of-the-art chips, maintaining these restrictions could unintentionally motivate China's technological advancements.[157] This scenario is exemplified by the launch of DeepSeek, which claims to achieve results comparable to leading U.S. models with significantly higher efficiency and lower costs. This development has sparked debate over the resource-heavy training methods employed by American AI firms like OpenAI, contributing to a staggering US$1 trillion decline in U.S. technology stocks. Meanwhile, there are suspicions that DeepSeek might have circumvented stringent U.S. export controls by using intermediaries in Singapore to acquire advanced Nvidia chips for training its model, prompting an investigation by the U.S. Department of Commerce.[158] Both Nvidia and Singapore's Ministry of Trade and Industry have refuted these allegations.[159]

As the AI race intensifies, the foreign trade policies of both countries increasingly target each other, leading to unpredictable outcomes. The most likely consequence is that these external constraints will decelerate AI development in both nations. However, such mutual exclusion might also compel each country toward greater self-reliance. For instance, U.S. export controls on semiconductor chips have catalyzed the advancement of China's domestic chipmaking, while Chinese export controls on raw materials have stimulated U.S. investments in critical minerals. Ultimately, whether these export restrictions will foster or hinder AI innovation in the long run remains an open question.

---

[155] Jacob Robbins, *China Outpaces US in Backing Chip Startups*, PitchBook (Jan. 25, 2024), https://pitchbook.com/news/articles/china-semiconductors-startups-vc.

[156] Liu Qianer, *China on Cusp of Next-Generation Chip Production Despite US Curbs*, Fin. Times (Feb. 6, 2024), https://www.ft.com/content/b5e0dba3-689f-4d0e-88f6-673ff4452977.

[157] Ariel Cohen, *China's Massive Barrage in the Chip Battle*, Forbes (May 31, 2024), https://www.forbes.com/sites/arielcohen/2024/05/31/chinas-massive-barrage-in-the-chip-battle/.

[158] Karen Freifeld, *US Looking into Whether DeepSeek Used Restricted AI Chips, Source Says*, Reuters (Feb. 1, 2025), https://www.reuters.com/technology/us-looking-into-whether-deepseek-used-restricted-ai-chips-source-says-2025-01-31/.

[159] *Press Statement on Whether DeepSeek Gained Access to US Export-Controlled Chips Through Intermediaries in Singapore*, Ministry of Trade & Indus. Sing. (Feb. 1, 2025), https://www.mti.gov.sg/Newsroom/Press-Releases/2025/02/Press-Statement-on-whether-DeepSeek-gained-access-to-US-export-controlled-chips.

## Conclusion

The legal framework for fostering technological innovation and economic growth presents a complex challenge. This chapter delves into a comparative analysis of the legal infrastructures in the United States and China, assessing their impacts on the AI competition between these two countries, particularly through the lenses of data privacy, copyright, and export restrictions.

In the realm of data privacy, Chinese companies collaborating with government agencies for surveillance purposes may benefit from unparalleled access to vast amounts of personal data. This access facilitates the training of their AI models, providing a significant competitive edge. In contrast, privacy reforms in the United States could potentially restrict similar data access for American AI firms, impacting their model training capabilities.

Regarding copyright issues, while the United States Copyright Office has recently acknowledged the copyrightability of human compilations of AI-generated contents, China has adopted a more favorable stance toward AI innovation. Unlike the United States, where courts and the Copyright Office have declined to grant copyright protection to AI-generated works, Chinese courts have recognized the intellectual contributions of programmers who develop algorithms and users who configure prompts. This progressive approach grants copyright protection to AI-generated works, potentially accelerating AI innovation in China.

Lastly, the United States has implemented export restrictions targeting China's semiconductor industry, aiming to block access to advanced semiconductor chips and thus impede China's AI development. The long-term effects of these restrictions on China's technological progress in semiconductor chip technologies remain uncertain. This strategic move highlights the intense rivalry and strategic maneuvering prevalent in the global AI race.

Overall, the legal, financial, and institutional landscapes in the United States and China reflect a distinct ideological divergence between the two countries. China's model of public-guided private investment offers notable benefits in terms of capital deployment, infrastructure development, and strategic development. In contrast, the United States boasts a more fluid capital market and a more agile commercialization mechanism, which together foster AI entrepreneurship. Additionally, the United States continues to hold an edge in attracting global AI talents. However, increasing restrictive policies within the United States[160] are gradually eroding these advantages. Ultimately, this geopolitical rift is likely to impede the critical goal of fostering inclusive innovation in the AI sector.

## Acknowledgments

This study was made possible by the Direct Research Grant (Project No. 4059076) at the Chinese University of Hong Kong Faculty of Law.

---

[160] An increasing number of restrictive measures are being introduced, spanning from exportation to investment to overseas students. Anniek Bao, *U.S. Blacklists over 50 Chinese Companies in Bid to Curb Beijing's AI, Chip Capabilities*, CNBC (Mar. 26, 2025), https://www.cnbc.com/2025/03/26/us-blacklists-50-chinese-companies-in-bid-to-curb-beijings-ai-chip-capabilities.html; Fu Ting, *Republican Legislation Seeks to Ban Chinese Nationals from Studying in the US*, AP NEWS (Mar. 15, 2025), https://apnews.com/article/chinese-student-republican-visa-ban-946e613a8dace3b0092580fa6fe4e50b.